\documentclass[10pt]{iopart}
 \usepackage[dvips]{graphicx}
 \expandafter\let\csname equation*\endcsname\relax      
 \expandafter\let\csname endequation*\endcsname\relax   
 \usepackage{amsmath}
 \usepackage{amsfonts}
 \usepackage{amssymb}
 \usepackage{ifthen}                 
 \usepackage[dvipsnames]{xcolor}
 \newcommand{\showlabel}[1]{
   \label{#1}
 }

 \renewcommand{\sf}{\sffamily}
 \renewcommand{\bf}{\bfseries}
 \renewcommand{\it}{\itshape}

 \allowdisplaybreaks[4]     

 \renewcommand{\d}{{\rm d}}     
 
 \newcommand{\pd}[2]{\frac{\partial #1}{\partial #2} }     

 \newcommand{\ol}[1]{\overline{#1}}
 
 \newcommand{\ot}[1]{\tilde{#1}}

 \newcommand{\lra}{\leftrightarrow}

 \newcommand{\nn}{\nonumber}
 \newcommand{\er}[1]{(\ref{#1})}          

 \renewcommand{\L}{Lema\^{\i}tre}
 \newcommand{\LT}{{\L}--Tolman}
 \newcommand{\Rt}{\dot{R}}

 \newcommand{\tlr}{\tilde r}
 \newcommand{\tP}{\tilde P}

\begin{document}

 \title{\bf \sf Reversing the Null Limit of the Szekeres Metric}

 \author{\sf
 Charles Hellaby$^1$
 and
 Otakar Sv\'{\i}tek$^2$
 }

 \address{$^1$ Dept of Maths and Applied Maths, University of Cape Town, Rondebosch, 7701, South Africa}

 \address{$^2$ Institute of Theoretical Physics, Faculty of Mathematics and Physics, Charles University, V Hole\v{s}ovi\v{c}k\'{a}ch 2, 180 00 Prague 8, Czech Republic}

 \eads{Charles.Hellaby@uct.ac.za and ota@matfyz.cz}

 \date{}


 \begin{abstract}
 The null limits of the {\LT} and Szekeres spacetimes are known to be the Vaidya and news-free Robinson--Trautman metrics.  We generalise this result to the case of non-zero $\Lambda$, and then ask whether the reverse process is possible --- is there a systematic procedure to retrieve the timelike-dust metric from the null-dust case?  We present such an algorithm for re-constructing both the metric and matter tensor components of the timelike-dust manifold.  This undertaking has elucidated the null limit process, highlighted which quantities approach unity or zero, and necessitated a careful discussion of how the functional dependencies are managed by the transformations and substitutions used.  
 \end{abstract}


 \maketitle

 \section{\sf Previous Work and Motivation}

 The null limit of the (LT-type) Szekeres metric%
 \footnote{It is shown in \cite{Hell96b} and section 19.6.3 of \cite{PleKra06} that the KS-type Szekeres metric is a well-behaved limit of the LT-type, so there is really only one type.}%
 \cite{Szek75a,Szek75b} was first found by Gleiser \cite{Glei84}, but only in the quasi-spherical case.  His derivation used a transformation similar to that below, introduced a limiting parameter ($\Lambda \to \infty$), and retained only leading terms in the calculation, making one or two assumptions about limiting behaviours, such as $\phi_{,x}$ (i.e. $R'$ below) being finite and non-zero.  He identified the resulting metric as a ``pure radiation'' form of the Robinson--Trautman (RT) metric \cite{RobTra62}.  This paper foreshadowed a number of the later results mentioned in this section, and deserved more attention than it got.
 
 Lemos \cite{Lemo92} showed how to obtain the Vaidya metric \cite{Vaid51,Vaid53} as the null limit of some cases of the {\LT} metric \cite{Lem33,Tol34}.  This was generalised by Hellaby \cite{Hell94b}, who showed that the limiting metric (in the outgoing form) must either stop radiating and settle down to a Schwarzschild vacuum, or radiate away all its mass leaving a Minkowski vacuum.

 Hellaby subsequently applied the same process to the Szekeres metric in \cite{Hell96b}, and identified the result as a generalisation of the Kinnersley Rocket \cite{Kinn62}%
 \footnote{Hellaby was not aware of Gleiser's paper at the time, and it would appear Lemos was not either.  All 3 null limit papers are mentioned in Krasinski's book \cite{Kra97}, Gleiser's in \S 2.4.2 on p 30, the other two in \S 3.6 on p 122.}%
 .  It is easy to see that the null metric in \cite{Hell96b}, with $\epsilon = +1$, is the same as that in \cite{Glei84}.

 The Szekeres metric \cite{Szek75a,Szek75b} is an exact inhomogeneous spacetime that solves the Einstein field equations (EFEs) for comoving dust.  It has no simple symmetries, though its spatial sections are conformally flat.  Notably it contains no gravitational radiation, a feature shared by the Kinnersley rocket.

 There is a related class of spacetimes, generally lacking Killing vectors, whose properties are determined by a null geodesic congruence, as opposed to a timelike one in the Szekeres spacetime.  Specifically, this null congruence is expanding, shearfree and twistfree, and the associated solutions belong to the Robinson--Trautman class \cite{RobTra60,RobTra62}.  This class contains the Schwarzschild, Vaidya and C-metric solutions and their generalisations.  Chru\'{s}ciel and Singleton showed that generic vacuum members of this class, that represent deformations of a Schwarzschild black hole, lose their non-sphericity exponentially fast by radiating gravitational waves, and thereby approach the Schwarzschild geometry asymptotically \cite{Chru1,Chru2,ChruSin}.  Similar behavior is observed when a null fluid is included, but the asymptotic geometry approaches that of the spherically-symmetric Vaidya solution, as shown by Bi\v{c}\'{a}k, Perj\'{e}s, Podolsk\'{y} and Sv\'{i}tek \cite{BicakPerjes:1987,PodSvi05}.  
Potential interest in this class of geometries containing exact gravitational waves and/or null fluids, which can represent an effective description of high-frequency gravitational waves, has been significantly enhanced due to the recent detection of gravitational waves by LIGO collaboration \cite{ligo}.

 In \cite{DaMoGl96}, Dain, Moreschi and Gleiser responded to the contemporaneous discussion about photon rockets and gravitational radiation, some of it using approximate methods, by pointing out that the RT metrics include exact solutions for generalised photon rockets, as well as cases that include gravitational radiation.  A similar point was made by Podolsk\'{y} in \cite{Podo08} in a paper that considered the more general RT metrics with non-zero $\Lambda$, generalised existing photon rockets to include $\Lambda$, and looked at their properties.

 Bi\v{c}\'{a}k and Kucha\v{r} \cite{BicKuc97} set up a Langrangian and Hamiltonian framework for treating a null dust source in GR.  They found that timelike dust metrics have an extra degree of freedom, the zero point of time along each worldline, which null dust metrics lack, because of the ambiguity of affine parametrisation along the null geodesics.

 The spherically symmetric null fluid model considered by Gair \cite{Gair02} is different, in that it contains angular momentum.  Each expanding (or contracting) spherical shell contains a superposition of randomly directed photons, all with the same angular momentum, and all remaining in that shell.  The random superposition ensures the spacetime has no net angular momentum.  To obtain the limit of zero angular momentum is not straightforward, but requires a singular transformation in order to show it becomes the Vaidya metric.  We will not consider such models here.

 An interesting question about the null limits of \cite{Glei84,Lemo92,Hell96b} noted above is whether one may do the reverse process --- can one find a procedure for creating a metric with a timelike fluid flow, starting from a null-fluid metric?  In what follows, we address this task.

 \section{\sf The Szekeres Metric}

 The Szekeres metric is
 \begin{align}
   \d s^2 = - \d t^2 + \frac{\left( R' - \dfrac{R E'}{E} \right)^2 \d r^2}{W^2}
      + \frac{R^2}{E^2} \big( \d p^2 + \d q^2 \big) ~,
      \showlabel{ds2Sz}
 \end{align}
 where $R = R(t,r)$, $W = \sqrt{\epsilon + f}\;$, $f = f(r)$ and $\epsilon = +1, 0, -1$.  
 We use a prime to denote the $r$-derivative, while an overdot is used for the $t$-derivative.
 The function $E$ is
 \begin{align}
   E = \frac{S}{2} \left( \frac{(p - P)^2}{S^2} + \frac{(q - Q)^2}{S^2} + \epsilon \right) ~,
   \showlabel{ErSPQ}
 \end{align}
 where $S = S(r)$, $P = P(r)$, $Q = Q(r)$.  
 From the above form for $E$, one may check that
 \begin{align}
   E E'_{pp} + E' E_{qq} - E_p E'_p - E_q E'_q & = 0   \showlabel{Erel1} \\
   E E'_{qq} + E' E_{pp} - E_p E'_p - E_q E'_q & = 0   \showlabel{Erel2} \\
   E_p^2 - E E_{pp} + E_q^2 - E E_{qq} + \epsilon & = 0   \showlabel{Erel3} \\
   E_{pq} & = 0   \showlabel{Erel4}
 \end{align}
 and these allow simplification when the metric form \er{ds2Sz} is inserted into the EFEs.  Given that the matter is comoving dust,
 \begin{align}
   T^{ab} = \rho u^a u^b ~,~~~~~~ u^a = \delta^a_t ~,   \showlabel{SzTabua}
 \end{align}
 the evolution equation and the density then follow:
 \begin{align}
   \Rt^2 & = \frac{2 M}{R} + f + \frac{\Lambda R^2}{3} ~,
      \showlabel{EvEq} \\
   \kappa \rho & = \frac{2 \left( M' - \dfrac{3 M E'}{E} \right)}{R^2 \left( R' - \dfrac{R E'}{E} \right)} ~,
      \showlabel{kDen}
 \end{align}
 where the arbitrary function $M = M(r)$ is a mass-like factor in the gravitational potential term $M/R$.

 Since we will be taking the limit as $f \to \infty$, we only consider $f > 0$ solutions.  With $\Lambda = 0$, \er{EvEq} has a parametric solution,
 \begin{align}
   R & = \frac{M}{f} (\cosh\eta - 1) ~,~~~~~~ (\sinh\eta - \eta) = \sigma \frac{f^{3/2}}{M} (t - a) ~,
   \showlabel{HypEv}
 \end{align}
 but if non-zero $\Lambda$ is included, then we can write the solution as a formal integral along worldlines of constant $(r, \theta, \phi)$,
 \begin{align}
   t - a = \int_0^R \frac{1}{\sigma} \left[ \dfrac{2 M}{\ot{R}} + f + \dfrac{\Lambda \ot{R}^2}{3} \right]^{-1/2} \, \d \ot{R} ~.
   \showlabel{tofRLam}
 \end{align}
 In \er{HypEv} and \er{tofRLam}, $a = a(r)$ is another arbitrary function that gives the bang time, the time on each constant $r$ worldline when $R = 0$, 
 and $\sigma=\pm 1$ admits time reversal.

 The Szekeres metric can be thought of as a non-symmetric generalisation of the \LT\ and Ellis metrics, \cite{Lem33,Tol34,Elli67}.  The arbitrary functions $f$, $M$ and $a$ determine the geometry and evolution of the underlying symmetric model, while the functions $S$, $P$ and $Q$ control the deviation from spherical, planar, or pseudo-spherical symmetry.  For a fuller description of the Szekeres metric, see \cite{Hel09,Kra97}.

 \section{\sf The Robinson--Trautman Metric}

The general form of the Robinson--Trautman spacetime including cosmological constant and null fluid (generally of Petrov type~II) can be given by the following line element \cite{RobTra60,RobTra62}
\begin{equation}
\d s^2 = -2H\,\d v^2-\,2\,\d v\,\d \tlr + \frac{\tlr^2}{\tP^2}\,(\d y^{2} + \d x^{2}),\showlabel{RTmetric}
\end{equation}
where ${2H = \Delta(\,\ln{\tP}) -2\tlr(\,\ln{\tP})_{,v} -{2M(v)/\tlr} -(\Lambda/3) \tlr^2}$,
\begin{equation}\showlabel{Laplace}
\Delta\equiv \tP^2(\partial_{xx}+\partial_{yy}),
\end{equation}
and $\Lambda$ is the cosmological constant. The metric depends on two functions, ${\,\tP(v,x,y)\,}$ and ${\,M(v)\,}$, which satisfy the nonlinear Robinson--Trautman equation
\begin{equation}
\Delta\Delta(\,\ln{\tP})+12\,M(\,\ln{\tP})_{,v}-4\,M_{,v}=2\kappa\, n^2\,.
\showlabel{RTequationgen}
\end{equation}
where the function $n(v,x,y)$ corresponds to the density of the null fluid, with energy momentum tensor ${T_{ab}=n^2(v,x,y)\,\tlr^{-2}\,k_{a} k_{b}}$, where ${\mathbf{k}=\partial_{\tlr}}$ is aligned along the degenerate principal null direction (we use the convention ${G_{ab}+\Lambda g_{ab}=\kappa\, T_{ab}}$).
The ``mass'' function $M(v)$ may be set to a constant by a suitable coordinate transformation for the vacuum solution ($n=0$). It is related to the Bondi mass in asymptotically flat cases.

The spacetime admits a geodesic, shearfree, twistfree and expanding null congruence, generated by ${\mathbf{k}=\partial_{\tlr}}$. The coordinate $\tlr$ is an affine parameter along this congruence, $v$~is a retarded time coordinate with the $v=$
constant hypersurfaces being null, and $x,y$ are spatial coordinates spanning the transversal 2-spaces with their Gaussian curvature (for ${\tlr=1}$) being given by
\begin{equation}
\mathcal{K}(v,x,y)\equiv\Delta(\,\ln{\tP})\,.
\showlabel{RTGausscurvature}
\end{equation}
For general fixed values of $\tlr$ and $v$, the Gaussian curvature is ${\mathcal{K}/\tlr^2}$ so that, as ${\tlr\to\infty}$, they become locally flat.

 Usually, it is assumed that the transversal 2-spaces are compact and simply-connected, which leads to the $S^{2}$ topology, and such solutions can then correspond to deformations of spherically-symmetric black holes. For a specific value of function $\tP$, namely
\begin{equation}
  \tP_0=1+\frac{1}{4}\left(x^2+y^2\right)
\end{equation}
we obtain $\mathcal{K}=1$ (consistent with spherical symmetry) and the metric reduces to the dynamical type D Vaidya(--(anti-)de Sitter) solution.  General Robinson--Trautman spacetimes of type II with $S^{2}$ topology (with  $\mathcal{K}$ no longer a constant) thus represent generalizations of these spherically symmetric geometries to a non-symmetric dynamical situation containing (exact) gravitational waves. This gravitational radiation facilitates their asymptotic transition to spherical symmetry.

  There are as well non-symmetric type D solutions with null radiation in Robinson--Trautman class which are generally known as Kinnersley rockets and have a non-constant Gaussian curvature $\mathcal{K}(v)$, but $\mathcal{K}=1$ can be achieved via coordinate transformation.

 Comparison of \er{ds2Sz} and \er{RTmetric}, together with the demand of preserving the interpretation of the transversal two-spaces spanned by $(p,q)$, resp.\ $(x,y)$, leads us to identify the roles played by functions $E$ and $\tP$ in the Szekeres and Robinson--Trautman spacetimes. This subsequently results in an expected correspondence between $R$ and $\tlr$ in the limit to be considered. The limit should also identify the direction of timelike dust flow $u^a$ with the null fluid flow direction $k^a$. All these considerations will be made explicit in the following sections.

 \section{\sf The Null Limit of Szekeres Dust}

 First, let us review how the null limit is obtained.  In fact the following is a slight generalisation of previous results, since we allow $\Lambda$ to be non-zero%
 \footnote{The $\Lambda$ in \cite{Glei84} is the limit parameter, not the cosmological constant.}%
 .

 In the evolution equation \er{EvEq}, the function $f$ is twice the energy per unit mass of the matter particles (as well determining the local spatial curvature in \er{ds2Sz}).  To obtain a null limit, in which the dust particles achieve light speed, we let the energy diverge, $f \to \infty$.  For a sensible metric, though, we require that $R$ and $M$ remain finite in the limit.  Consequently, we need to transform the metric before taking the limit, so that the limit process is well behaved.

 By \er{EvEq} and \er{HypEv}, finite $R$ and $M$ means
 \begin{align}
   \Rt^2 - f = \frac{2 M}{R} & + \dfrac{\Lambda R^2}{3} ~~~~\mbox{remains finite in the limit}
      \showlabel{Rt2fLim} \\
   \Rt^2 - W^2 = \frac{2 M}{R} & + \dfrac{\Lambda R^2}{3} - \epsilon ~~~~\mbox{remains finite in the limit}
      \showlabel{Rt2W2Lim} \\
   \Lambda = 0:~~~~~~~~ \cosh\eta - 1 \sim f & ~~~~~~\to~~~~~~
      \cosh\eta \sim \sinh\eta \sim \frac{e^\eta}{2} \sim f
      \showlabel{f-eta} \\
   (t - a) = \frac{M (\sinh\eta - \eta)}{\sigma f^{3/2}} & \sim \frac{M (\cosh\eta - 1)}{\sigma f^{3/2}}
      \sim \frac{R f}{\sigma f^{3/2}} = \frac{R}{\sigma \sqrt{f}\;}
      \showlabel{ta-eta-f} \\
   \Lambda \neq 0:~~~~~~~~~~~~~~~~~~~~~ t - a & \sim \int_0^R \frac{1}{\sigma \sqrt{f}\;} \, \d \ot{R} = \frac{R}{\sigma \sqrt{f}\;} ~.
      \showlabel{ta-eta-fLam}
 \end{align}
 From \er{f-eta} it follows that $|\eta|\to\infty$ for $f\to\infty$, which implies late time evolution, i.e. far from the bang.  However \er{ta-eta-f} or \er{ta-eta-fLam} shows that $(t-a)\to 0$, and this is because proper time ceases to run as the particles approach light speed: the null limit generates infinite time dilation.

 Differentiating \er{tofRLam} with respect to $r$, we have
 \begin{align}
   - a' & = \pd{}{r} \int_0^R \frac{1}{\sigma} \left[ \dfrac{2 M}{\ot{R}} + f + \dfrac{\Lambda \ot{R}^2}{3} \right]^{-1/2}
         \, \d \ot{R} \\
   & = \frac{R'}{\sigma} \left[ \dfrac{2 M}{R} + f + \dfrac{\Lambda R^2}{3} \right]^{-1/2} \nn \\
      &~~~~ - \int_0^R \frac{1}{\sigma} \left( \frac{M'}{\ot{R}} + \frac{f'}{2} \right)
         \left[ \dfrac{2 M}{\ot{R}} + f + \dfrac{\Lambda \ot{R}^2}{3} \right]^{-3/2} \, \d \ot{R} ~, \\
   \to~~~~~~~~ R' & = \left[ \dfrac{2 M}{R} + f + \dfrac{\Lambda R^2}{3} \right]^{1/2} \Bigg\{ - \sigma a' \nn \\
      &~~~~ + \int_0^R \left( \frac{M'}{\ot{R}} + \frac{f'}{2} \right)
         \left[ \dfrac{2 M}{\ot{R}} + f + \dfrac{\Lambda \ot{R}^2}{3} \right]^{-3/2} \, \d \ot{R} \Bigg\} ~,
 \end{align}
 and thus, as $f \to \infty$, the limiting forms for $R$, $\Rt$, $R'$, $W^2$ are
 \begin{align}
   W^2 & \to f   \showlabel{limW2} \\
   R & \to \sigma \sqrt{f}\; (t - a)   \showlabel{limR} \\
   \Rt & \to \sigma \sqrt{f}\;   \showlabel{limRt} \\
   R' & \to U f   \showlabel{limRr} \\
   \mbox{where}~~~~~~~~ U & = \frac{R f'}{2 f^2} - \frac{\sigma a'}{\sqrt{f}\;} ~.   \showlabel{UDef}
 \end{align}
 We note that the limiting value of $R'$ is ambiguous.  How does $f'$ behave in the $f \to \infty$ limit?  If we choose $f' = 0$, and $a'$ happens to be zero, is $R' = 0$ acceptable?  Should we insist $a' \ne 0$?  The ambiguity of the limiting behaviour of $R'$ and $U$ is at least partly due to the re-scaling freedom of the $r$ coordinate.  An important step below allows us to avoid knowing how $f'$ behaves.

 We now apply the following transformation:
 \begin{align}
   R & = R(t,r) ~, & v & = \frac{R}{2 f \sigma} + \int_0^r \frac{a'}{\sqrt{f}\;} \, \d r ~,
   \showlabel{RtrvtrDef}
 \end{align}
 and from \er{RtrvtrDef}, it is clear that the $R$-dependence of $v$ disappears as $f \to \infty$, meaning $v \to v(r)$.  It also seems that we require $a'/\sqrt{f}\,$ to be finite and non-zero, so as to ensure $v$ does not become degenerate.  However, $a' = 0$ would be acceptable at individual points.  As noted in \cite{Hell94b,Hell96b}, a regular pseudo-spherical origin requires $f \to 0$, which is not compatible with $f \to \infty$, so the behaviour of $a'$ at an origin is not relevant.  For this transformation, the Jacobi matrix and its inverse:
 \begin{align}
   J & = \pd{(v, R)}{(t, r)} =
   \begin{pmatrix}
      \dfrac{\Rt}{2 f \sigma} & \dfrac{R' - 2 f U}{2 f \sigma} \\[4mm]
      \Rt & R'
   \end{pmatrix}
   , &
   J^{-1} & = \pd{(t, r)}{(v, R)} =
   \begin{pmatrix}
      \dfrac{R' \sigma}{\Rt U} & - \dfrac{R' - 2 f U}{2 f \Rt U} \\[4mm]
      - \dfrac{\sigma}{U} & \dfrac{1}{2 f U}
   \end{pmatrix}
   ,
   \showlabel{JcbMx}
 \end{align}
 have limiting forms
 \begin{align}
   J & = \pd{(v, R)}{(t, r)} = \begin{pmatrix}
      \dfrac{1}{2 \sqrt{f}\;} & \dfrac{- U}{2 \sigma} \\[4mm]
      \sigma \sqrt{f}\; & f U
   \end{pmatrix}
   ~,~~~~~~ &
   J^{-1} & = \pd{(t, r)}{(v, R)} = \begin{pmatrix}
      \sqrt{f}\; & \dfrac{1}{2 \sigma \sqrt{f}\;} \\[4mm]
      - \dfrac{\sigma}{U} & \dfrac{1}{2 f U}
   \end{pmatrix}
   ~,
   \showlabel{JcbMx0}
 \end{align}
 and the inverse transformation is:
 \begin{align}
   \d t & = \frac{R' \sigma}{\Rt U} \, \d v - \frac{(R' - 2 f U)}{2 f \Rt U} \, \d R
      \showlabel{dtdvdR} ~, \\
   \d r & = - \dfrac{\sigma}{U} \, \d v + \dfrac{1}{2 f U} \, \d R ~.
      \showlabel{drdvdR}
 \end{align}
 Inserting the transformations \er{dtdvdR} and \er{drdvdR} into the Szekeres metric \er{ds2Sz} gives
 \begin{align}
   \d s^2 & = - \left( \frac{R' \sigma}{\Rt U} \, \d v - \frac{(R' - 2 f U)}{2 f \Rt U} \, \d R \right)^2
         + \frac{\left( R' - \dfrac{R E'}{E} \right)^2}{W^2}
            \left( - \dfrac{\sigma}{U} \, \d v + \dfrac{1}{2 f U} \, \d R \right)^2 \nn \\
      &~~~~~~ + \frac{R^2}{E^2} \big( \d p^2 + \d q^2 \big) \nn \\
   & = \left( \frac{R^2 (E'/E)^2}{W^2 U^2}
         - \frac{2 R R' (E'/E)}{W^2 U^2}
         + \frac{R'^2 \big( \Rt^2 - W^2 \big)}{\Rt^2 W^2 U^2} \right) \d v ^2 \nn \\
      &~~~~ - \sigma \left( \frac{R^2 (E'/E)^2}{W^2 f U^2} - \frac{2 R R' (E'/E)}{W^2 f U^2} + \frac{2 R'}{\Rt^2 U}
         + \frac{R'^2 \big( \Rt^2 - W^2 \big)}{\Rt^2 W^2 f U^2} \right) \d v \, \d R \nn \\
      &~~~~ + \left( \frac{R^2 (E'/E)^2}{4 W^2 f^2 U^2} - \frac{R R' (E'/E)}{2 W^2 f^2 U^2}
         + \frac{R'^2 \big( \Rt^2 - W^2 \big)}{4 \Rt^2 W^2 f^2 U^2} + \frac{(R' - f U)}{\Rt^2 f U} \right) \d R^2 \nn \\
      &~~~~ + \frac{R^2}{E^2} \big( \d p^2 + \d q^2 \big) ~.
      \showlabel{ds2T1}
 \end{align}
 From the evolution equation \er{EvEq} and the $W$ definition we have,
 \begin{align}
   \Rt^2 - W^2 & =
      \frac{2 M}{R} + \frac{\Lambda R^2}{3} - \epsilon \equiv B ~,
      \showlabel{Rt2W2}
 \end{align}
 which we now substitute into \er{ds2T1}.  This is a step that is hard to reverse, because $R$ becomes a coordinate in the null metric, and there is no evolution equation for it.

 Now, from \er{JcbMx} the transformation of $E'$ would be
 \begin{align}
   \left. \pd{E}{v} \right|_R & =
      \left. \pd{E}{r} \right|_t \left. \pd{r}{v} \right|_R + \left. \pd{E}{t} \right|_r \left. \pd{t}{v} \right|_R
      = E' \left( - \dfrac{\sigma}{U} \right) + 0 \left( \dfrac{R' \sigma}{\Rt U} \right) \\
   \to~~~~~~~~ \left. \pd{E}{v} \right|_R & = - \dfrac{\sigma E'}{U} ~~~~\lra~~~~ E' = - \sigma U \left. \pd{E}{v} \right|_R ~.
   \showlabel{EvEr}
 \end{align}
 However, it is important for a well-defined limit that we do not use this.  Instead we write
 \begin{align}
   E^* & = - \dfrac{\sigma f E'}{R'} ~~~~\lra~~~~ E' = - \dfrac{\sigma R'E^*}{f} ~.
   \showlabel{Estar}
 \end{align}
 Nevertheless, by \er{limRr}, $R' \to U f$ in the null limit, so comparing \er{EvEr} and \er{Estar}, this implies that $E^* \to E_v$ in that same limit.

 Putting \er{Rt2W2} and \er{Estar} into \er{ds2T1} leads to
 \begin{align}
   & \d s^2 = \nn \\
   & \left( \frac{R^2 (- \sigma R')^2 (E^*/E)^2}{(f)^2 W^2 U^2}
         - \frac{2 R R' (- \sigma R') (E^*/E)}{(f) W^2 U^2}
         + \frac{2 \sigma R R'^2 (E^*/E)}{W^2 f U^2}
         + \frac{R'^2 \big( B \big)}{\Rt^2 W^2 U^2} \right) \d v^2 \nn \\
      &~ - \sigma \left( \frac{R^2 (- \sigma R')^2 (E^*/E)^2}{(f)^2 W^2 f U^2}
         - \frac{2 R R' (- \sigma R') (E^*/E)}{(f) W^2 f U^2}
         + \frac{2 R'}{\Rt^2 U}
         + \frac{R'^2 \big( B \big)}{\Rt^2 W^2 f U^2} \right) \d v \, \d R \nn \\
      &~ + \left( \frac{R^2 (- \sigma R')^2 (E^*/E)^2}{4 (f)^2 W^2 f^2 U^2}
         - \frac{R R' (- \sigma R') (E^*/E)}{2 (f) W^2 f^2 U^2}
         + \frac{R'^2 \big( B \big)}{4 \Rt^2 W^2 f^2 U^2} + \frac{(R' - f U)}{\Rt^2 f U} \right) \d R^2 \nn \\
      &~ + \frac{R^2}{E^2} \big( \d p^2 + \d q^2 \big) \\
   & = \left( \frac{R^2 R'^2 (E^*/E)^2}{W^2 f^2 U^2}
         + \frac{2 \sigma R R'^2 (E^*/E)}{W^2 f U^2}
         + \frac{R'^2 B}{\Rt^2 W^2 U^2} \right) \d v^2 \nn \\
      &~~~~ - \sigma \left( \frac{R^2 R'^2 (E^*/E)^2}{W^2 f^3 U^2}
         + \frac{2 \sigma R R'^2 (E^*/E)}{W^2 f^2 U^2}
         + \frac{2 R'}{\Rt^2 U}
         + \frac{R'^2 B}{\Rt^2 W^2 f U^2} \right) \d v \, \d R \nn \\
      &~~~~ + \left( \frac{R^2 R'^2 (E^*/E)^2}{4 W^2 f^4 U^2}
         + \frac{\sigma R R'^2 (E^*/E)}{2 W^2 f^3 U^2}
         + \frac{R'^2 B}{4 \Rt^2 W^2 f^2 U^2}
         + \frac{(R' - f U)}{\Rt^2 f U} \right) \d R^2 \nn \\
      &~~~~ + \frac{R^2}{E^2} \big( \d p^2 + \d q^2 \big) ~.
         \showlabel{ds2T2}
 \end{align}
 We are now ready to take the limit of the line element.  First, applying \er{limW2}-\er{UDef} reduces the above to
 \begin{align}
   & \d s^2 = \nn \\
      & \left( \frac{R^2 (f U)^2 (E^*/E)^2}{(f) f^2 U^2}
         + \frac{2 \sigma R (f U)^2 (E^*/E)}{(f) f U^2}
         + \frac{(f U)^2 B}{(f) (f) U^2} \right) \d v^2 \nn \\
      &~~ - \sigma \left( \frac{R^2 (f U)^2 (E^*/E)^2}{(f) f^3 U^2}
         + \frac{2 \sigma R (f U)^2 (E^*/E)}{(f) f^2 U^2}
         + \frac{2 (f U)}{(f) U}
         + \frac{(f U)^2 B}{(f) (f) f U^2} \right) \d v \, \d R \nn \\
      &~~ + \left( \frac{R^2 (f U)^2 (E^*/E)^2}{4 (f) f^4 U^2}
         + \frac{\sigma R (f U)^2 (E^*/E)}{2 (f) f^3 U^2}
         + \frac{(f U)^2 B}{4 (f) (f) f^2 U^2}
         + \frac{((f U) - f U)}{(f) f U} \right) \d R^2 \nn \\
      &~~ + \frac{R^2}{E^2} \big( \d p^2 + \d q^2 \big) \\
   & = \left( \frac{R^2 (E^*/E)^2}{f}
         + 2 \sigma R (E^*/E)
         + B \right) \d v^2 \nn \\
      &~~~~ - \sigma \left( \frac{R^2 (E^*/E)^2}{f^2}
         + \frac{2 \sigma R (E^*/E)}{f}
         + 2
         + \frac{B}{f} \right) \d v \, \d R \nn \\
      &~~~~ + \left( \frac{R^2 (E^*/E)^2}{4 f^3}
         + \frac{\sigma R (E^*/E)}{2 f^2}
         + \frac{B}{4 f^2}
         + \frac{0}{f} \right) \d R^2 \nn \\
      &~~~~ + \frac{R^2}{E^2} \big( \d p^2 + \d q^2 \big) ~,
 \end{align}
 and second, letting $f \to \infty$ removes several of these terms.  
 As noted above, \er{limRr}, \er{Estar} and \er{EvEr}, show the limit of $E^*$ is
 \begin{align}
   E^* & = - \dfrac{\sigma f E'}{R'} ~~\to~~ - \dfrac{\sigma E'}{U} = E_v ~,
   \showlabel{Ev-form}
 \end{align}
 and in the limit, $v$ becomes a function of $r$ only, as is evident from the transformation \er{RtrvtrDef}.

 The relations \er{Erel1}-\er{Erel4} carry over, with $r$ derivatives being replaced by $v$ derivatives, and \er{ErSPQ} is retained, except that $S$, $P$ and $Q$ become functions of $v$, not $r$.  In the reverse process, controlling the functional dependence of $E$ is possibly the trickiest aspect.

 The limiting metric then is
 \begin{align}
   \d s^2 & = - \left\{ \epsilon - \frac{2 M}{R} - \frac{\Lambda R^2}{3} - 2 \sigma \frac{R E_v}{E} \right\} \d v^2
      - 2 \sigma \, \d v \, \d R + \frac{R^2}{E^2} \big( \d p^2 + \d q^2 \big)
      \showlabel{ds2NS}
 \end{align}
 Here two functions have vanished: $f(r)$ and also $a(r)$, and there is no evolution DE for $\Rt$.  However we still have $M(v)$, $S(v)$, $P(v)$ and $Q(v)$.  Comparing the above metric with \er{RTmetric} we find that the identification
 \begin{align}
   \sigma = +1 ~,~~~~~~ p \to x ~,~~~~~~ q \to y ~,~~~~~~ R \to \tlr ~,~~~~~~ \tP = E
   \showlabel{SzRTid}
 \end{align}
 leads to 
 \begin{align}
   \mathcal{K} \equiv \Delta(\ln{\tP}) = \epsilon ~,~~~~~~
   2 H = \epsilon - \frac{2 M}{R} - \frac{\Lambda R^2}{3} - 2 \sigma \frac{R E_v}{E} ~,
 \end{align} 
 and therefore \er{ds2NS} is the RT subclass corresponding to constant Gaussian curvature, \er{RTGausscurvature}.  This subclass has no gravitational radiation, and is the only subclass with a zero gravity-wave news function \cite{Bonn94,Corn00,Ivan05}.

 \subsection{\sf The Limit of the Matter Tensor}

 As with $E'$ in \er{Estar}, we also replace $M'$ according to
 \begin{align}
   M' & = - \dfrac{\sigma R' M^*}{f} \ \showlabel{Mr-form} ~,
 \end{align}
 and transform the Szekeres matter tensor into
 \begin{align}
   \kappa T^{ab} & = \frac{2 \big( M' - 3 M (E'/E) \big)}{R^2 \big( R' - R (E'/E) \big)}
      \begin{pmatrix}
         \frac{\Rt^2}{4 f^2} & \frac{\sigma \Rt^2}{2 f} & 0 & 0 \\[1mm]
         \frac{\sigma \Rt^2}{2 f} & \Rt^2 & 0 & 0 \\[1mm]
         0 & 0 & 0 & 0 \\
         0 & 0 & 0 & 0
      \end{pmatrix}
      \nn \\
   & = \frac{2 \big( (- \sigma R'/f) M^* - 3 M (- \sigma R'/f)(E^*/E) \big)}{R^2 \big( R' - R (- \sigma R'/f)(E^*/E) \big)}
      \begin{pmatrix}
         \frac{\Rt^2}{4 f^2} & \frac{\sigma \Rt^2}{2 f} & 0 & 0 \\[1mm]
         \frac{\sigma \Rt^2}{2 f} & \Rt^2 & 0 & 0 \\[1mm]
         0 & 0 & 0 & 0 \\
         0 & 0 & 0 & 0
      \end{pmatrix}
      \nn \\
   & = \frac{2 \big( M^* - 3 M (E^*/E) \big)}{R^2 \big( - \sigma f - R (E^*/E) \big)}
      \begin{pmatrix}
         \frac{\Rt^2}{4 f^2} & \frac{\sigma \Rt^2}{2 f} & 0 & 0 \\[1mm]
         \frac{\sigma \Rt^2}{2 f} & \Rt^2 & 0 & 0 \\[1mm]
         0 & 0 & 0 & 0 \\
         0 & 0 & 0 & 0
      \end{pmatrix}
      ~,
      \nn
 \intertext{then, using \er{limW2}-\er{UDef} and taking the null limit, we arrive at}
   \kappa T^{ab} & \sim \frac{2 \big( M^* - 3 M (E^*/E) \big)}{R^2}
      \begin{pmatrix}
         \frac{- \sigma}{4 f^2} & \frac{- 1}{2 f} & 0 & 0 \\[1mm]
         \frac{- 1}{2 f} & - \sigma & 0 & 0 \\[1mm]
         0 & 0 & 0 & 0 \\
         0 & 0 & 0 & 0
      \end{pmatrix}
      \nn \\
   \to~~~~~~~~ \kappa T^{ab} & = \frac{2 \big( M^* - 3 M (E^*/E) \big)}{R^2}
      \begin{pmatrix}
         0 & 0 & 0 & 0 \\[1mm]
         0 & - \sigma & 0 & 0 \\[1mm]
         0 & 0 & 0 & 0 \\
         0 & 0 & 0 & 0
      \end{pmatrix}
      = \kappa \ol{\rho} \, k^a k^b ~.
      \showlabel{TabNS}
 \end{align}
 We find $M^* \to M_v$ for exactly the same reason that $E^* \to E_v$.  
 Inserting the metric \er{ds2NS} into the EFEs gives the same result.  To ensure positive energy density, if the radiation is outgoing we must have, $M^* \leq0$ with $\sigma = +1$, and vice versa.

 The limiting matter tensor \er{TabNS} agrees with the RT expression \er{RTequationgen}, since the identification \er{SzRTid} gives us
 \begin{align}
   \Delta \Delta (\ln{\tP}) = 0 ~,~~~~~~
   \frac{E_v}{E} \to \big( \ln \tP)_{,v} ~,~~~~~~
   \ol{\rho} \to \frac{n^2}{\tlr^2} ~.
 \end{align}
 There is a curious point here.  In the full RT metric, eq \er{RTequationgen} is a dynamical equation, since $v$ is the only time-varying coordinate.  However, in our limit-transformation, the $v$ of the null metric \er{RTmetric} replaces the $r$ of the timelike metric \er{ds2Sz}, which is a spatial coordinate, suggesting that \er{RTequationgen}, in the constant curvature case, is the null-limit version of \er{kDen}, the Szekeres density equation.

 \section{\sf Going Backwards}
 \showlabel{GoBack}

 The task at hand is, given the null-fluid metric \er{ds2NS}, is there a procedure for constructing or retrieving an associated timelike-fluid metric from it?

 The most obvious approach to undoing the limiting process just described, is to take the steps in reverse order and attempt to undo each one.  However, the first problem is how to re-introduce the terms and functions that vanished in the limiting process.  We try to make this step algorithmic by re-introducing terms based on: what is already there, the coordinate transformation, and maybe the form of the target metric.  Probably the EFEs will be needed to obtain all functions and evolution equations.

 Starting with \er{ds2NS}, we put it in the form
 \begin{align}
   \d s^2 & = \big( 2 \sigma A + B \big) \d v^2
      - 2 \sigma \big( 1 \big) \d v \, \d R
      + \big( 0 \big) \d R^2
      + \frac{R^2}{E^2} \big( \d p^2 + \d q^2 \big) ~,
 \end{align}
 where $A$ contains the factor $E_v/E$.

 We shall be applying the transformation
 \begin{align}
   R = R(t,r) ~~~~\to~~~~ \d R & = \Rt \, \d t + R' \, \d r ~,
      \showlabel{Rtr} \\
   v = v(t,r) ~~~\mbox{such that}~~~ \d v & = \left( \frac{\sigma \Rt}{2 f} \right) \d t
      + \sigma R' \left( \frac{1}{2 f} - \frac{1}{\beta} \right) \d r ~,
      \showlabel{vtr}
 \end{align}
 where $f = f(r)$ and $\beta$ is unspecified.  
 All introduced parameters are functions of $(t, r)$ unless stated otherwise; these parameters are: $\beta$, $\chi$, $\psi$; the only exception will be the function $h = h(t,r,p,q)$ and later $k = k(t,r,p,q)$.  
 The Jacobi matrices of \er{Rtr} and \er{vtr} are
 \begin{align}
   J^{-1} = \pd{(t,r)}{(v,R)} = 
      \begin{pmatrix}
         \frac{\beta}{\sigma \Rt} & \frac{(2 f - \beta)}{2 f \Rt} & 0 & 0 \\[1mm]
         \frac{- \beta}{\sigma R'} & \frac{\beta}{2 f R'} & 0 & 0 \\[1mm]
         0 & 0 & 1 & 0 \\
         0 & 0 & 0 & 1
      \end{pmatrix}
   ,~
   J = \pd{(v,R)}{(t,r)} = 
      \begin{pmatrix}
         \frac{\sigma \Rt}{2 f} & \frac{\sigma R' (\beta - 2 f)}{2 f \beta} & 0 & 0 \\[1mm]
         \Rt & R' & 0 & 0 \\[1mm]
         0 & 0 & 1 & 0 \\
         0 & 0 & 0 & 1
      \end{pmatrix}
   .
   \showlabel{JBack}
 \end{align}
 But before applying the transformation, we need to modify the metric as follows.  First, insert some factors --- $\frac{h \beta^2}{f W^2}$, $\frac{\beta^2}{\Rt^2 W^2}$, $\chi$ --- which stand for quantities that went to unity when the null limit was taken,
 \begin{align}
   \d s^2 & = \left( 2 \sigma A \frac{h}{f} + B \frac{1}{\Rt^2} \right) \left( \frac{\beta^2}{W^2} \right) \, \d v^2
      - 2 \sigma \big( \chi \big) \d v \, \d R
      + \big( 0 \big) \d R^2
      + \frac{R^2}{E^2} \big( \d p^2 + \d q^2 \big) ~,
      \showlabel{ds2InFctrs}
 \intertext{where $W^2 = \epsilon + f$.  Second, add some terms --- $\frac{A^2 h^2 \beta^2}{f^2 W^2}$, $\psi$ --- quantities that vanished in the null limit,}
   \d s^2 & = \left( A^2 \frac{h^2}{f^2} + 2 \sigma A \frac{h}{f} + B \frac{1}{\Rt^2} \right)
      \left( \frac{\beta^2}{W^2} \right) \, \d v^2
      - 2 \sigma \big( \chi \big) \d v \, \d R
      + \big( \psi \big) \d R^2 \nn \\
   &~~~~ + \frac{R^2}{E^2} \big( \d p^2 + \d q^2 \big) ~.
      \showlabel{ds2Two}
 \intertext{Third, copy the $A$- and $B$-terms from $g_{vv}$ to $g_{vR}$ and $g_{RR}$ multiplied by powers of $(\sigma/2 f)$}
   \d s^2 & = \left( A^2 \frac{h^2 \beta^2}{W^2 f^2}
         + 2 \sigma A \frac{h \beta^2}{W^2 f}
         + B \frac{\beta^2}{W^2 \Rt^2} \right) \d v^2 \nn \\
      &~~~~ - 2 \sigma \left( A^2 \frac{h^2 \beta^2}{2 W^2 f^3}
         + 2 \sigma A \frac{h \beta^2}{2 W^2 f^2}
         + B \frac{\beta^2}{2 W^2 \Rt^2 f}
         + \chi \right) \d v \, \d R \nn \\
      &~~~~ + \left( A^2 \frac{h^2 \beta^2}{4 W^2 f^4}
         + 2 \sigma A \frac{h \beta^2}{4 W^2 f^3}
         + B \frac{\beta^2}{4 W^2 \Rt^2 f^2}
         + \psi \right) \d R^2 \nn \\
      &~~~~ + \frac{R^2}{E^2} \big( \d p^2 + \d q^2 \big) ~.
      \showlabel{ds2Three}
 \end{align}
 Comparing this with \er{ds2T2}, it is evident that $\beta$ plays the role of $R'/U$, but the introduction of the extra factor of $h$ turns out to be rather important, as demonstrated below.  We could have chosen $\chi = \beta/\Rt^2$ and $\psi = (\beta - f)/(\Rt^2 f)$ from the start, but instead we will obtain these below from the required form of the metric.

 Now we apply the transformation \er{Rtr}-\er{vtr}, which converts the above metric to
 \begin{align}
   \d s^2 & = - \frac{\Rt^2 (\chi - \psi f)}{f} \, \d t^2
         + \frac{2 \Rt R' (\chi f - \chi \beta + \psi \beta f)}{\beta f} \, \d r \, \d t \nn \\
      &~~~~ + \Bigg( A^2 \frac{h^2 R'^2}{W^2 f^2}
         + 2 \sigma A \frac{h R'^2}{W^2 f}
         + B \frac{R'^2}{W^2 \Rt^2} 
         + \frac{R'^2 (2 \chi f - \chi \beta + \psi \beta f)}{\beta f} \Bigg) \, \d r^2 \nn \\
      &~~~~ + \frac{R^2}{E^2} \big( \d p^2 + \d q^2 \big) ~.
 \end{align}
 If we specify that $g_{tt} = -1$ and $g_{tr} = 0$, we find
 \begin{align}
   - \frac{\Rt^2 (\chi - \psi f)}{f} & = -1 ~~~~\to~~~~
      \chi - \psi f = \frac{f}{\Rt^2} \\
   \frac{2 \Rt R' (\chi f - \beta (\chi - \psi f))}{\beta f} & = 0 = 
      \frac{2 \Rt R' (\chi f - \beta(f/\Rt^2))}{\beta f} \\
      \to~~~~~~~~ \chi & = \frac{\beta}{\Rt^2} ~,~~~~~~ \psi = \frac{\beta - f}{\Rt^2 f} ~.
 \end{align}
 Hence
 \begin{align}
   \d s^2 & = - \d t^2 + \Bigg( A^2 \frac{h^2 R'^2}{W^2 f^2} + 2 \sigma A \frac{h R'^2}{W^2 f} + B \frac{R'^2}{W^2 \Rt^2} 
      + \frac{\Rt^2}{R'^2} \Bigg) \, \d r^2 + \frac{R^2}{E^2} \big( \d p^2 + \d q^2 \big) \nn \\
   & = - \d t^2 + \Bigg( \left( \frac{h A}{f} + \sigma \right)^2 \frac{R'^2}{W^2}
      + \big( B + W^2 - \Rt^2 \big) \frac{R'^2}{W^2 \Rt^2} \Bigg) \, \d r^2
      + \frac{R^2}{E^2} \big( \d p^2 + \d q^2 \big) ~,
 \end{align}
 and restoring what $A$ and $B$ are, we get
 \begin{align}\label{metric-Ev}
   \d s^2 & = - \d t^2 + \Bigg( \left( \frac{h R (E_v/E)}{f} + \sigma \right)^2 \frac{R'^2}{W^2} \nn \\
      &~~~~ + \big( 2 M/R + \Lambda R^2/3 - \epsilon + W^2 - \Rt^2 \big) \frac{R'^2}{W^2 \Rt^2} \Bigg) \, \d r^2
         + \frac{R^2}{E^2} \big( \d p^2 + \d q^2 \big) ~.
 \end{align}
 Expressing $E_v$ using a newly defined function $E^\varkappa$
 \begin{align}
   E_v = \frac{- \sigma f E^\varkappa}{h R'} ~,
   \showlabel{EvEkB}
 \end{align}
 and substituting into \er{metric-Ev} gives
 \begin{align}
   \d s^2 & = - \d t^2 + \Bigg( \left( \frac{- \sigma h f R (E^\varkappa/E)}{h R' f}
         + \sigma \right)^2 \frac{R'^2}{W^2} \nn \\
      &~~~~ + \big( 2 M/R + \Lambda R^2/3 - \epsilon + W^2 - \Rt^2 \big) \frac{R'^2}{W^2 \Rt^2} \Bigg) \, \d r^2
         + \frac{R^2}{E^2} \big( \d p^2 + \d q^2 \big) \nn \\
   & = - \d t^2 + \Bigg( \big( R' - R (E^\varkappa/E) \big)^2 \frac{1}{W^2} \nn \\
      &~~~~ + \big( 2 M/R + \Lambda R^2/3 - \epsilon + W^2 - \Rt^2 \big) \frac{R'^2}{W^2 \Rt^2} \Bigg) \, \d r^2
         + \frac{R^2}{E^2} \big( \d p^2 + \d q^2 \big) ~.
      \showlabel{FinalMetric}
 \end{align}
 We notice that the factors $\beta$ and $h$ have vanished.

 Because the transformation \er{vtr} gives $v$ in the form $v(t, r)$, it is not immediately obvious that the $E^\varkappa$ and $E$ defined via \er{EvEkB} is independent of $t$.  This is where $h$ comes in --- we are free to specify that $h(t,r,p,q)$ is such that the new $E^\varkappa/E$ does not depend on $t$.  The justification of \er{EvEkB} is discussed in section \ref{RstFnFm}, where it is shown how a function of $v$ only can become a function of $r$ only.  (Below we will introduce $M^\varkappa$ and require that it too must depend on $r$ only.)  Therefore, since $E$ does not depend on $t$, the relations \er{Erel1}-\er{Erel4} are retained, because they held in $v$-form in \er{ds2NS}.

 We do not have an evolution equation with which to eliminate the second term in $g_{rr}$.  Thus we use the EFEs (assuming $E$ does not depend on $t$), and requiring $G^p{}_r = 0 = G^q{}_r$, we get (without needing \er{Erel1}-\er{Erel4}) for example
 \begin{align}
   G^p{}_r & = \frac{R'^2 E^2 (E E'_p - E' E_p) (2 M/R + \Lambda R^2/3 - \epsilon + W^2 - \Rt^2)}
   {R (\Rt^2 R^2 E' (R E' - 2 R' E) + R R'^2 E^2 (2 M/R + \Lambda R^2/3 + W^2 - \epsilon))} = 0
 \end{align}
 which recovers the evolution equation \er{EvEq}, and puts the metric into the Szekeres form, meaning it is a silent, irrotational dust spacetime.

 \subsection{\sf Undoing the Limit of the Matter Tensor}
 \showlabel{UnLimTab}

 We begin with the null metric matter tensor \er{TabNS},
 \begin{align}
   T^{ab} & = \frac{2 \big( M_v - 3 M (E_v/E) \big)}{R^2}
      \begin{pmatrix}
         0 & 0 & 0 & 0 \\[1mm]
         0 & - \sigma & 0 & 0 \\[1mm]
         0 & 0 & 0 & 0 \\
         0 & 0 & 0 & 0
      \end{pmatrix}
      ~,
 \end{align}
 and we insert a factor in the common denominator, 
 \begin{align}
   T^{ab} & = \frac{2 \big( M_v - 3 M (E_v/E) \big)}{R^2 \left( \dfrac{f}{k \Rt^2} \right)}
      \begin{pmatrix}
         0 & 0 & 0 & 0 \\[1mm]
         0 & - \sigma & 0 & 0 \\[1mm]
         0 & 0 & 0 & 0 \\
         0 & 0 & 0 & 0
      \end{pmatrix}
      ~.
 \end{align}
 Here we have introduced the function $k(t, r, p, q)$, which serves the same purpose as the $h$ introduced in eq \er{ds2InFctrs}.  The manner in which the correct functional dependence is restored is discussed in section \ref{RstFnFm}.  We next add in a vanishing term
 \begin{align}
   T^{ab} & = \frac{2 \big( M_v - 3 M (E_v/E) \big)}{R^2 \left( \dfrac{f}{k \Rt^2} + \dfrac{\sigma R}{\Rt^2} (E_v/E)  \right)}
      \begin{pmatrix}
         0 & 0 & 0 & 0 \\[1mm]
         0 & - \sigma & 0 & 0 \\[1mm]
         0 & 0 & 0 & 0 \\
         0 & 0 & 0 & 0
      \end{pmatrix}
      ~,
 \end{align}
 and we copy the $T^{RR}$ component into the $T^{vR}$ and $T^{vv}$ components, divided by factors of $2 f/\sigma$,
 \begin{align}
   T^{ab} & = \frac{2 \big( M_v - 3 M (E_v/E) \big)}{R^2 \left( \dfrac{f}{k \Rt^2} + \dfrac{\sigma R}{\Rt^2} (E_v/E)  \right)}
      \begin{pmatrix}
         \frac{-\sigma}{4 f^2} & \frac{-1}{2 f} & 0 & 0 \\[1mm]
         \frac{-1}{2 f} & - \sigma & 0 & 0 \\[1mm]
         0 & 0 & 0 & 0 \\
         0 & 0 & 0 & 0
      \end{pmatrix}
      ~.
 \end{align}
 This copying of terms could be written%
 \footnote{It does not seem to be possible to do this for the metric in going from \er{ds2Two} to \er{ds2Three}.}
 \begin{align}
   T^{cd} & \to V^c_a T^{ab} V^d_b ~,~~~~~~ \mbox{where}~~~~
   V^d_b = 
      \begin{pmatrix}
         1 & \frac{\sigma}{2 f} & 0 & 0 \\
         \frac{\sigma}{2 f} & 1 & 0 & 0 \\
         0 & 0 & 1 & 0 \\
         0 & 0 & 0 & 1
      \end{pmatrix}
      ~.
 \end{align}
 In fact, the values of the first and second terms in the top row of $V^d_b$ do not actually matter, but the above choice has the neat property that $V^d_b$ becomes the identity in the null limit.

 Next we substitute for $E_v$ and $M_v$ using \er{EvEkB} and
 \begin{align}
   M_v = \frac{- \sigma f M^\varkappa}{k R'}
   \showlabel{MvMkB}
 \end{align}
 giving
 \begin{align}
   T^{ab} & = \frac{2 \big( (- \sigma f M^\varkappa/k R') - 3 M (- \sigma f E^\varkappa/k R')/E \big)}
                   {R^2 \left( \dfrac{f}{k \Rt^2} + \dfrac{\sigma R}{\Rt^2} (- \sigma f E^\varkappa/k R')/E  \right)}
      \begin{pmatrix}
         \frac{-\sigma}{4 f^2} & \frac{-1}{2 f} & 0 & 0 \\[1mm]
         \frac{-1}{2 f} & - \sigma & 0 & 0 \\[1mm]
         0 & 0 & 0 & 0 \\
         0 & 0 & 0 & 0
      \end{pmatrix} \\
   & = \frac{2 \big( M^\varkappa - 3 M (E^\varkappa/E) \big)}{R^2 \Big( R' - R (E^\varkappa/E)  \Big)} (- \sigma \Rt^2)
      \begin{pmatrix}
         \frac{-\sigma}{4 f^2} & \frac{-1}{2 f} & 0 & 0 \\[1mm]
         \frac{-1}{2 f} & - \sigma & 0 & 0 \\[1mm]
         0 & 0 & 0 & 0 \\
         0 & 0 & 0 & 0
      \end{pmatrix}
      ~.
 \end{align}
 Finally we apply the inverse transformations \er{Rtr}-\er{vtr}, for which the Jacobi matrices are
 \begin{align}
   J^{-1} = \pd{(t,r)}{(v,R)} = 
      \begin{pmatrix}
         \frac{\beta}{\sigma \Rt} & \frac{(2 f - \beta)}{2 f \Rt} & 0 & 0 \\[1mm]
         \frac{- \beta}{\sigma R'} & \frac{\beta}{2 f R'} & 0 & 0 \\[1mm]
         0 & 0 & 1 & 0 \\
         0 & 0 & 0 & 1
      \end{pmatrix}
   ,\;
   J = \pd{(v,R)}{(t,r)} = 
      \begin{pmatrix}
         \frac{\sigma \Rt}{2 f} & \frac{\sigma R' (\beta - 2 f)}{2 f \beta} & 0 & 0 \\[1mm]
         \Rt & R' & 0 & 0 \\[1mm]
         0 & 0 & 1 & 0 \\
         0 & 0 & 0 & 1
      \end{pmatrix}
      ,
 \end{align}
 to get a form that is clearly $T^{ab} = \rho u^a u^b$, i.e. \er{SzTabua}, with $\rho$ the same as \er{kDen}:
 \begin{align}
   T^{ab} & = \frac{2 \big( M^\varkappa - 3 M (E^\varkappa/E) \big)}{R^2 \Big( R' - R (E^\varkappa/E)  \Big)} (- \sigma \Rt^2)
      \begin{pmatrix}
         \frac{-\sigma}{\Rt^2} & 0 & 0 & 0 \\[1mm]
         0 & 0 & 0 & 0 \\[1mm]
         0 & 0 & 0 & 0 \\
         0 & 0 & 0 & 0
      \end{pmatrix}
      ~.
      \showlabel{FinalTab}
 \end{align}

 \subsection{\sf Restoration of the Correct Functional Dependence}
 \showlabel{RstFnFm}

 A central problem with going backwards is that we are not taking a limit, so no terms disappear, meaning the final functional dependencies must be exact.  Indeed, terms have to be introduced, in order that the end result is correct.  This forces us to carefully examine the terms that disappear in the $f \to \infty$ limit.

 The transformations \er{Rtr}-\er{vtr} (or \er{RtrvtrDef}) do not allow a function of $r$ only to become a function of $v$ only, or vice-versa, so it is necessary to see how the procedure specified here makes this possible.  Firstly, when going forwards from $(t, r)$ to $(v, R)$, we have
 \begin{align}
   M_v & = \dot{M} \frac{\beta}{\sigma \Rt} + M' \left( \frac{-\beta}{\sigma R'} \right) \\
   M_R & = \dot{M} \frac{(2 f - \beta)}{2 f \Rt} + M' \frac{\beta}{2 f R'} ~.
 \end{align}
 For a function of $r$ only, with $\dot{M} = 0$, this gives
 \begin{align}
   M_R = \frac{- \sigma M_v}{2 f}
   \showlabel{MRMv}
 \end{align}
 which is true point by point.  Clearly, if $M_v$ is finite, then $M_R$ vanishes in the $f \to \infty$ limit.
 Conversely, for the backwards transformation,
 \begin{align}
   \dot{M} & = M_v \frac{\sigma \Rt}{2 f} + M_R \Rt \\
   M' & = M_v \frac{\sigma R' (\beta - 2 f)}{2 f \beta} + M_R R' ~,
 \end{align}
 both $M'$ and $\dot{M}$ are non-zero in general, unless we re-introduce $M_R$ by insisting that \er{MRMv} hold, in which case we get 
 \begin{align}
   \dot{M} & = M_v \frac{\sigma \Rt}{2 f} + \left( \frac{- \sigma M_v}{2 f} \right) \Rt = 0 \\
   M' & = M_v \frac{\sigma R' (\beta - 2 f)}{2 f \beta} + \left( \frac{- \sigma M_v}{2 f} \right) R'
      = \frac{- \sigma M_v R'}{\beta} ~.
 \end{align}
 This justifies \er{Mr-form}.  Of course, the re-introduction of $M_R$ implies a change in functional dependence from $M(v)$ to $M(v,R)$, which is the reason for the function $k$ used in \er{MvMkB}.  A similar argument justifies \er{EvEkB} and the function $h$, which converts $E(v, p, q)$ to $E(v, R, p, q)$ and then $E(r, p, q)$.

 \section{\sf Conclusion}

 The possibility that one may take a spacetime filled with moving dust (zero pressure matter), and allow the velocity of the dust particles to reach light speed, while still retaining a well-behaved spacetime, is a surprising and intriguing result, that is well worth understanding as thoroughly as we can.  While the process of taking this null limit has been known for some time, this paper is the first to ask whether the reverse process can be described in a systematic manner.  

 Firstly, in reviewing the process of taking the null limit of the Szekeres metric, we generalised previous results to the case of non-zero $\Lambda$, and arrived at the zero-news Robinson--Trautman--de Sitter metric.  Before taking the null limit, it is important to first transform the metric so that it does not become degenerate, and quantities that need to remain finite are preserved.  In addition, the spacelike ``radial'' coordinate $r$, that is comoving with the dust, must be replaced by the null coordinate $v$ that labels outgoing (or incoming) light paths.  Furthermore, it is important that the functions $E$ and $M$ are replaced by quantities that are not actually their transformations.  We also noted which steps would be challenging to reverse.  In particular, the null limit makes certain terms go to unity and others to zero, and certain functions lose their dependence on one or more variables.  The evolution equation \er{EvEq} does not survive, being used up in the transformation and the determination of limiting behaviours.

 Therefore, the backwards process involves re-introducing factors or functions that are ``initially'' one or nought, but subsequently aquire different values and functional dependencies.  In order to get these functional dependencies correct, two auxiliary functions $h$ and $k$ had to be introduced, even though they are re-absorbed before the process is finished.  The values of some of the introduced functions were determined by the expected form of the timelike metric, but the final evolution equation had to be derived from the Einstein equations.  This is because (i) the dynamical equation of the null metric maps to the density expression of the timelike one, and (ii) the null limit metric does not have an equation involving $\tilde{r}$ that would transform back into an evolution equation for $R(t,r)$ in the timelike metric.  The relation between the coordinate transformations, the limiting process, and the functional dependencies was clarified in section \ref{RstFnFm}.  

 It is possible there is a more streamlined approach to undoing the null limit, but it is clear that one needs some knowledge of the expected result in order to proceed.  Nevertheless, we have gained a more precise understanding of the forward null limit process along the way, since much more attention had to be paid to the limiting values of the terms, and how functional dependencies changed.

 The process of reversing the limit has highlighted another issue pertinent to limiting procedures applied to spacetime metrics, and that is its uniqueness.  As already stressed by Geroch \cite{geroch} such limits are coordinate dependent.  However, we have been led by a clear physical motivation regarding the matter content of the spacetime which substantially restricts influence of potential coordinate changes.  Different spacetime limits are distinguished \cite{geroch} according to the associated families of frames --- continuous sequences of frames changing with the parameter in which the limit is made, and having a well-defined frame in the limit.  Since we have preserved the transversal spaces (spanned by $p,q$ in the Szekeres geometry and $x,y$ in the Robinson--Trautman geometry) during the limiting process, they provide two well defined frame vectors (up to trivial 2-dimensional transformation) for the family.  Additionally, the direction of the fluid flow (changing from timelike $\partial_{t}$ in Szekeres to null $\partial_{\tilde{r}}$ in Robinson--Trautman) provides a third frame vector for the family of frames.  So the procedure substantially fixes the freedom in the limiting process when following geometrical and physical properties. According to \cite{geroch}, there are certain properties of spacetimes, called hereditary, that are preserved in the limit.  One such property is for example the absence of closed timelike curves.  Although the algebraic type of the Weyl tensor is not hereditary, the limit spacetime can be either of the same type or more special.  This means that the limit of the Szekeres spacetime, which is type $D$, cannot be of more general type.  It follows that in order to arrive at the most general type $II$ Robinson--Trautman spacetime, one would have to start from a type $II$ family of dust-containing cosmological spacetimes.

 The question of whether there are other metrics describing timelike or null dust filled spacetimes, that might be investigated in a similar manner is an interesting topic to pursue, 
 which could even lead to new metrics, or new insights into old ones.

 \end{document}